\documentclass[preprint]{mn2e}

\usepackage{epsfig}

\newcommand{\bmain}{\begin{list}{${\bullet}$}{}}
\newcommand{\emain}{\end{list}}
\newcommand{\bminor}{\begin{list}{$\bf\triangleright$}{}}
\newcommand{\eminor}{\end{list}}
\newcommand{\Ms}{\mbox{$\,\mbox{M}_\odot$}}
\newcommand{\Rs}{\mbox{$\,\mbox{R}_\odot$}}
\newcommand{\Ls}{\mbox{$\,\mbox{L}_\odot$}}
\newcommand{\kms}{\mbox{\,km\,s$^{-1}$}}

\newcommand{\Msyr}{\Ms\,\mbox{yr$^{-1}$}}

\title[The Evolution of a Surviving Companion after a Type Ia Supernova]
{On the Evolution and Appearance of a Surviving
Companion after a Type Ia Supernova Explosion}

\author[Ph.~Podsiadlowski]
{Ph.~Podsiadlowski\thanks{E-mail: podsi@astro.ox.ac.uk}\\
\it
University of Oxford, Department of Astrophysics, Oxford,
OX1 3RH
}

\date{\today}

\volume{000}

\pagerange{\pageref{firstpage}--\pageref{lastpage}} \pubyear{2003}

\begin{document}
\maketitle

\label{firstpage}

\begin{abstract}
One promising method to test progenitor models for Type Ia supernovae
is to identify surviving companion stars in historical supernova
remnants. A surviving companion will have been strongly affected by
its interaction with the supernova ejecta. Here we systematically
explore the evolution and appearance of a typical companion star that
has been stripped and heated by the supernova interaction during the
post-impact re-equilibration phase. We show that, depending on the
amount of heating and the amount of mass stripped by the impact or the
previous binary mass transfer, such a star may be significantly
overluminous or underluminous $10^3\,$--$\,10^4\,$yr after the
supernova relative to its pre-supernova luminosity (by up to two
orders of magnitude) and discuss the implications of these results for
the strategies to be employed in searches for such companions.
\end{abstract}


\section{Introduction}

In recent years, Type Ia supernovae (SNe Ia) have been used
successfully as cosmological probes of the Universe (Riess et al.\
1998; Perlmutter et al.\ 1999). However, the nature of their
progenitors has remained somewhat of a mystery. There is almost
universal agreement that they represent the disruption of a degenerate
object; but this is also where the agreement ends. There are numerous
progenitor models (for detailed reviews see, e.g., Branch et al.\
1995; Ruiz-Lapuente, Canal \& Isern 1997), but most of these have
serious theoretical/observational problems or do not appear to occur
in sufficient numbers to explain the observed frequency of SNe Ia in
our Galaxy ($\sim 3\times 10^{-3}\,$yr$^{-1}$; Cappellaro \& Turatto
1997).

The progenitor models can roughly be divided into three classes:
double-degenerate (DD) models, sub-Chandrasekhar models and
single-degenerate Chandrasekhar models. The double-degenerate model
(Iben \& Tutukov 1984; Webbink 1984) involves the merger of two CO
white dwarfs with a combined mass in excess of the Chandrasekhar mass
($\sim 1.4\Ms$). While this model has the advantage of being quite
common (see, e.g., Iben \& Tutukov 1986; Yungelson et al.\ 1994; Han
et al.\ 1995; Iben, Tutukov \& Yungelson 1997; Han 1998; Nelemans et
al.\ 2001), it seems more likely that the disruption of the lighter
white dwarf and the accretion of its debris onto the more massive one
leads to the transformation of the surviving CO white dwarf into an
ONeMg white dwarf which subsequently collapses to form a neutron star
(i.e.\ undergoes accretion-induced collapse; AIC) rather than
experience a thermonuclear explosion (Nomoto \& Iben 1985; Saio \&
Nomoto 1985, 1998; Timmes, Woosley \& Taam 1994; Mochkovitch, Guerrero
\& Segretain 1997).  There may be a small parameter range where AIC
can be avoided, but it is unlikely to account for more than a small
number of SNe Ia.

Woosley \& Weaver (1995) speculated that a sub-Chandrasekhar-mass white
dwarf could produce a SN Ia if helium was ignited violently in a shell
surrounding the CO core and triggered a detonation wave that propagated
inwards and ignited the CO core (see also Livne \& Arnett 1995). While
it is not clear whether this can even work in principle, simulations
of such explosions show that they do not reproduce the properties
of observed SNe Ia (see, e.g., the discussion in Wheeler 1996).

The arguably most favoured class of models at the present time
involves single-degenerate scenarios, where the white dwarf accretes
from a non-degenerate companion star (Whelan \& Iben 1973; Nomoto
1982).  In these models, the companion star can be in different
evolutionary phases and may either be a hydrogen-rich star or a helium
star. One of the major problems with these models is that it is
generally difficult to increase the mass of a white dwarf by accretion
due to the occurrence of nova explosions and/or helium flashes (Nomoto
1982) which may eject most of the accreted mass. There is a narrow
parameter range where a white dwarf can accrete hydrogen-rich material
and burn it in a stable manner, but this requires rather special
circumstances.

One promising channel that has been identified in recent years relates
them to supersoft X-ray sources (van den Heuvel et al.\ 1992;
Rappaport, Di\,Stefano \& Smith 1994; Li \& van den Heuvel 1997). In
this channel, the companion star is a somewhat evolved main-sequence
star or subgiant of 2\,--\,3\,\Ms, transferring mass on a thermal
timescale to a white dwarf. In Figure~1 we present a typical binary
calculation that illustrates this channel (also see Langer et al.\
2000; Han \& Podsiadlowski 2003). The initial system consists of a
2.1\Ms\ somewhat evolved main-sequence star and a 0.8\Ms\ white
dwarf. In this calculation we adopted the estimates for the accretion
efficiency of Hachisu, Kato \& Nomoto (1999) and assumed that any
excess matter that could not be acrreted by the white dwarf was
ejected from the system. As the calculation demonstrates, the white
dwarf is able to grow very effectively. When it reaches the
Chandrasekhar mass, it has parameters very similar to U Scorpii, a
supersoft binary and recurrent nova where the white dwarf is already
close to the Chandrasekhar mass (Hachisu et al.\ 2000; Thoroughgood et
al.\ 2001).

%%
% Figure 1: a supersoft model to produce U Sco
%%
%
\begin{figure}
\centerline{\psfig{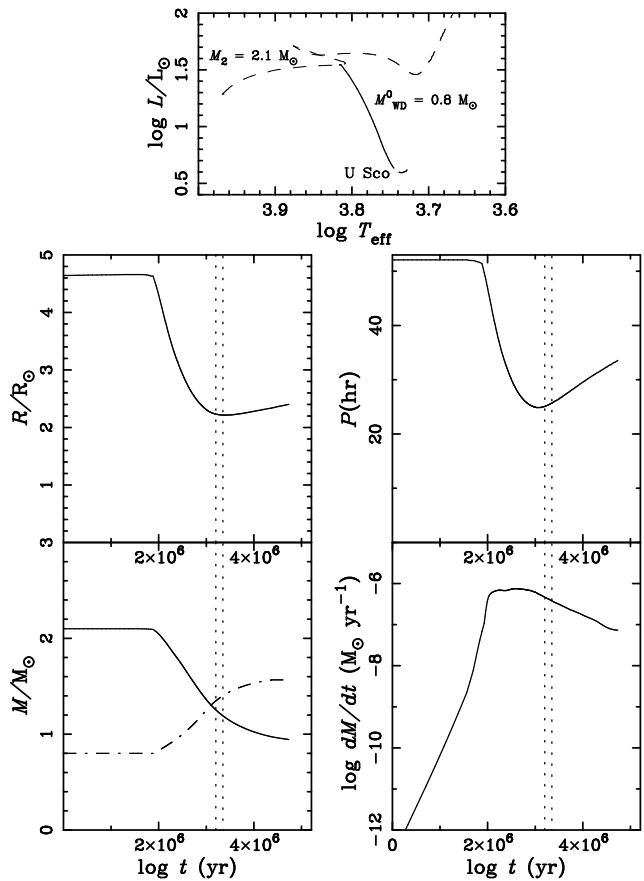}}
\caption{Evolutionary model for a SN Ia progenitor in the supersoft
channel (modelling a system similar to U Sco).  H-R diagram (top
panel) and key binary parameters as a function of time since the
beginning of mass transfer (with arbitrary offset). {\it Middle left:}
Radius of the secondary. {\it Middle right:} Orbital period. {\it
Bottom left:} Mass of the secondary (solid curve) and the white dwarf
(dot-dashed curve).  {\it Bottom right:} Mass-loss rate of the
secondary.  The secondary initially has a mass of 2.1\Ms\ and has
already exhausted most of its hydrogen in the core.  The white dwarf
has an initial mass of 0.8\Ms\ and is able to accrete according to the
formalism given by Hachisu et al.\ (1999).  The dashed curve in the
top panel shows the evolutionary track of a single 2.1\Ms\ star, the
solid curve the evolutionary track of the secondary during the
mass-transfer phase. The phase when the white dwarf is close to the
Chandrasekhar mass ($\sim 1.4\Ms$) is indicated in all panels: as a
gap in the track in the top panel; by vertical, dashed lines the lower
panels.}\label{fig1}
\end{figure}

While U Scorpii provides an excellent candidate for a SN Ia
progenitor, consistent with theoretical expectations, it would be even
better to have a more direct observational test for progenitor models
(e.g. Ruiz-Lapuente 1997), in particular since it is quite possible --
perhaps even likely -- that there is more than one channel that leads
to a SN Ia. Such direct tests could involve the detection of hydrogen
or helium in the ejecta or the supernova environment, which could come
from the outer layers of the exploding object, circumstellar material
that was ejected from the progenitor system (e.g. Cumming et al.\
1996), or matter that was stripped from the secondary by the supernova
interaction and was mixed into the ejecta. A particularly conclusive
test would be the detection of a companion star that has survived the
supernova explosion in the supernova remnant. At present the detection
of a surviving companion would only be feasible in our Galaxy, e.g. in
the supernova remnants SN 1006 and Tycho (Ruiz-Lapuente et al.\ 2002).

A surviving companion can be identified by its expected unusual
properties (Marietta, Burrows \& Fryxell 2000; Canal, M\'endez \&
Ruiz-Lapuente 2001): (1) it will be a runaway star moving directly
away from the centre of the supernova remnant, where the runaway
velocity will be similar to the orbital velocity of the companion star in
the pre-supernova binary (e.g. $\sim 170\kms$ for a system like U Sco;
$\sim 20\kms$ for a symbiotic binary with a giant companion); (2)
the secondary will have been strongly affected by its interaction with
the supernova ejecta and have an unusual appearance; and (3)
the composition of the secondary may show chemical anomalies
and be enriched with heavy elements produced in the thermonuclear
explosion (similar to what is observed in Nova Sco, a black-hole
binary where the companion star has been enriched with supernova
material; Israelian et al.\ 1999; Podsiadlowski et al.\ 2002).

Marietta et al.\ (2000) have recently performed the most detailed and
most systematic hydrodynamical investigation of the impact of the
supernova shell in a SN Ia with hydrogen-rich secondaries of various
types, representing different progenitor channels (for earlier studies
see Wheeler, Lecar \& McKee 1975; Fryxell \& Arnett 1981; Taam \&
Fryxell 1984; Livne, Tuchman \& Wheeler 1992). In particular, they
showed that main-sequence and subgiant companions will lose $\sim
15\,$per cent of their mass due to the supernova impact, while giant
companions will be stripped of almost all of their hydrogen-rich
envelopes. Furthermore, the interior of the secondaries will be
significantly shock-heated by the supernova impact; as a consequence
the immediate post-impact luminosity will be increased dramatically.

It is the purpose of this paper to model the further evolution of a
subgiant secondary to clarify the principles that govern their
post-impact evolution and to determine their expected appearance which
will be useful for searches of surviving companions in historical SN
Ia remnants which are presently under way. In Section~2 we
systematically examine the consequences of heating and mass stripping
for the evolution and appearance of the secondary and in Section~3 we
discuss the implications of these results.

\section{Post-supernova heating and stripping}

Immediately after the impact with the supernova ejecta, the secondary
will be strongly out of thermal equilibrium and is likely to be
significantly puffed up and overluminous (Marietta et al.\ [2000]
estimate that a subgiant may have a luminosity as high as $\sim
5000\Ls$).  The subsequent evolution is governed by how the star
re-establishes thermal equilibrium, which depends on the amount of
heating and mass loss it has experienced. Because these effects have
very different consequences for the evolution of the star, the
secondary may appear overluminous {\em or} underluminous at different
times during the equilibration phase, as we will demonstrate below.

For this study we employed a standard Henyey-type stellar evolution
code, described in detail in Podsiadlowski, Rappaport \& Pfahl (2002),
which we have previously used to study the effects of external
irradiation and tidal heating in stars (Podsiadlowski 1991,
1996). Since we cannot follow the dynamical interaction of the
secondary during the impact phase, we adopt the following procedure to
model the effects of mass stripping and supernova heating.  We first
perform a mass-loss calculation where we take off mass from the star
at a very high rate (typically $10^{-2}\Msyr$), so that mass loss can
be considered essentially adiabatic (except in the outermost
layers). To model the residual heating of the envelope (i.e. the extra
heating of the part of the envelope that was not stripped off in the
supernova impact), we add a constant and uniform heating source to the
remaining outer layers of the star with a total luminosity of
$10^{5}\Ls$ for a certain of time and then follow the subsequent
re-equilibration of the star. This procedure is of course only very
approximate, but should nevertheless be sufficiently realistic to
study the basic physical aspects of the problem, the main objective of
this paper\footnote{We have tested that the subsequent evolution of
the star is not very sensitive to the timescale on which the initial
stripping and heating is simulated (as long as it is short compared to
the thermal timescale of the outer layers of the star); while the
initial appearance may differ substantially for different
implementations, the evolutionary tracks and internal structures tend
to converge rapidly in the subsequent re-equilibration phase.}.

%%
% Figure 2: Heating calculations for a subgiant
%%
%
\begin{figure}
\centerline{\psfig{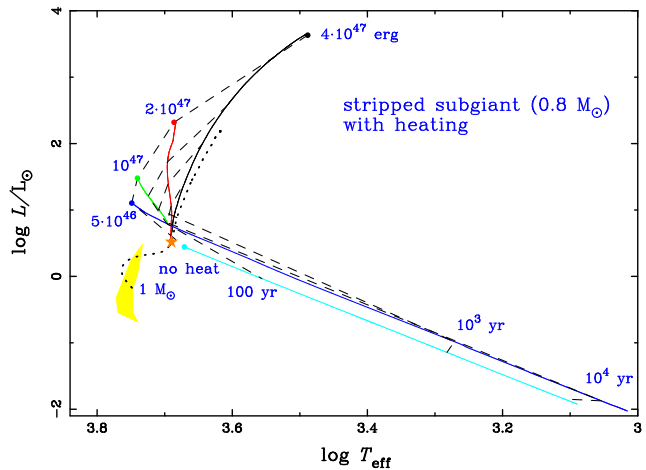}}
\caption{Evolutionary tracks (solid curves) in the H-R diagram of a
1\Ms\ subgiant that has been stripped of the outermost 0.2\Ms\ of its
envelope and has been heated by various amounts of energy by the
supernova impact. The amount of energy deposited in the outermost
90\,per cent of the star (by radius) is indicated at the beginning of
each track. The dashed curves give from left to right the location of
the stars at the beginning and $10^2$, $10^3$ and $10^4$\,yr after the
beginning of each sequence, respectively. The dotted curve shows the
evolutionary track of a 1\Ms\ star, and the star symbol indicates the
position of the initial, undisturbed model used for the supernova
impact simulations. For comparison, the lightly shaded (yellow) region
shows the position of a subgiant that had been stripped of 0.8\Ms\ of
its envelope (i.e.\ only has a total mass of 0.2\Ms\ after the impact)
between $10^3$ and $10^4\,$yr after the supernova.}
\label{fig2}
\end{figure}

In Figure~2 we present evolutionary tracks in the Hertzsprung-Russell
(H-R) diagram illustrating the post-impact equilibration phase for a
subgiant (with an initial pre-supernova mass of 1\Ms) that was
stripped of the outermost 0.2\Ms\ of its envelope by the supernova
impact and was heated by various different amounts of energy (as
indicated at the beginning of each track). The initial subgiant has a
radius of 2.5\Rs\ and is very similar to the model of the secondary
shown in Figure~1 at the time of the supernova.  In these simulations,
the energy was deposited uniformly in the outermost 90\,per cent of
the radial extent of the star which contained 0.57\Ms. The energy in
the calculation with the largest amount of energy deposited ($4\times
10^{47}\,$erg) is close to the energy needed to unbind these layers
completely, which we estimate to be $\sim 4.9\times 10^{47}\,$erg. As
a consequence, the star is initially very puffed up with a radius of
230\,\Rs\ and has a luminosity of 4300\Ls\ (cf.\ Marietta et al.\
2000). Because of the large radius and the large luminosity, the
thermal timescale of the puffed-up envelope is very short initially,
and the star contracts very rapidly along a Hayashi track. After
$10^3$\,yr its luminosity has dropped to 200\Ls\ and after $10^4$\,yr
to 27\Ls. In this context we emphasize that in all the models shown in
Figure~2 the evolution is initially much faster than the
Kelvin-Helmholtz timescale, $t_{\rm KH}$, of the pre-supernova
subgiant ($\sim 10^6\,$yr in our case), since the timescale on which
the star evolves in the H-R diagram is determined by the thermal
timescale of the rapidly evolving outer layers of the star, which can
be many orders of magnitude shorter than $t_{\rm KH}$.  (Of course, it
will take a full Kelvin-Helmholtz time before the star has
re-established thermal equilibrium completely.)

In the calculations with $10^{47}$ and $2\times 10^{47}\,$erg, the
evolution is similar although less dramatic. On the other hand, in the
case where the star is only heated by $5\times 10^{46}\,$erg or is not
heated at all, the evolution is dramatically different, and the
secondary becomes significantly underluminous during the early thermal
re-equilibration phase (with a luminosity of $\sim 10^{-1}\Ls$ after
$10^3\,$yr and $\sim 10^{-2}\Ls$ after $10^4\,$yr).  This behaviour is
similar to the behaviour commonly found in binary mass-transfer
calculations where a secondary strongly out of thermal equilibrium can
be very underluminous (e.g.\ Langer et al.\ 2000; King et al.\ 2001;
Podsiadlowski et al.\ 2002).

%%
% Figure 3: Internal structure for star being stripped (without heating)
%%
%
\begin{figure}
\centerline{\psfig{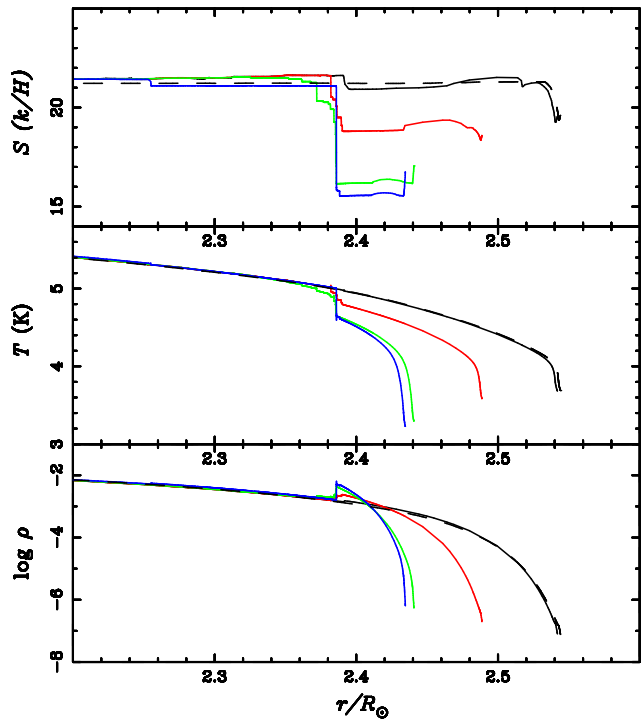}}
\caption{Evolution of the thermodynamic properties during the thermal
equilibration phase for a 1\Ms\ subgiant that has been stripped of the
outermost 0.2\Ms\ by the impact with a supernova (without supernova
heating). {\it Top:} entropy, {\it middle:} temperature, {\it bottom:}
density, all as a function of radius for the outermost part of the
star. The solid curves show the profiles at different times after the
stripping (from top to bottom/right to left: immediately after the
stripping, after $\sim 100\,$yr, $\sim 10^3\,$yr, $\sim
10^4\,$yr). The dashed curves show the final thermal equilibrium
structure.}
\label{fig3}
\end{figure}

The principal reason for this behaviour is that, in a subgiant whose
outer layers have been suddenly stripped off, the layers in the deep
interior of the envelope have to expand significantly since the final
thermal equilibrium radius is similar to the initial radius. The
energy needed to drive this expansion comes from the nuclear
luminosity produced in the core of the star, reducing the radiative
luminosity that reaches the outer parts of the star. In the case where
the star is dramatically out of thermal equilibrium, as is the case
here, essentially all the nuclear luminosity is used up in the
expansion of the envelope. As a consequence, the very outer layers of
the star are no longer re-supplied with energy from below; but because
of the continued radiative losses from the surface, the outer layers
start to cool dramatically.

To illustrate this behaviour, Figure~3 shows the early evolution of
the thermodynamic structure in the outer parts of the envelope for the
stripped subgiant in Figure~2 that has not been heated (the region
shown contains about 1\,per cent of the mass of the star).  As the
outer layers cool (particularly clearly seen in the drop of the
entropy), this outer layer contracts, and the star in fact becomes
somewhat undersized relative to its final equilibrium radius (it takes
a full Kelvin-Helmholtz time, $\sim 10^6\,$yr in the case shown, for
the star to re-establish thermal equilibrium). During this cooling
phase, the star evolves roughly along a line of constant radius in the
H-R diagram and would have the appearance of an anomalously red
subgiant. Since the surface luminosity becomes much smaller than
the luminosity inside the star, only a small fluctuation of the
luminosity profile (e.g. due the generally non-linear behaviour of the
equilibration process which can drive luminosity waves through the
star) may lead to a sudden change of the surface luminosity, and we
therefore suspect that such a star would be highly
variable\footnote{In our calculations we regularly observe sudden
changes of the luminosity profile, where the star suddenly brightens
by an order of magnitude or more in a single time step and
subsequently continues to cool. However, at present this behaviour is
not well resolved numerically.}.

Because our models were constructed in a very different fashion from
the models of Marietta et al.\ (2000), it is not entirely clear which
of our sequences most closely resembles their subgiant case. The fact
that the subgiant is stripped of a significant fraction ($\sim
15\,$per cent) of its mass certainly implies that the remaining
envelope will also be strongly heated.  Comparing the radii of our
models to the radius ($\sim 9\Rs$) found by Marietta et al.\ (2000)
then suggests that the sequences with $10^{47}$ and $2\times
10^{47}\,$erg of heating are probably the most comparable.  In these
sequences, the secondary resembles a typical subgiant with a
luminosity of 10\,--25\Ls\ after $10^3\,$yr and 6\,--\,10\Ls\ after
$10^4\,$yr, still significantly overluminous relative the initial
model which had a luminosity of $\sim 3\Ls$.

The timescale on which the surface luminosity evolves after the
supernova impact is determined mainly by the thermal timescale of the
envelope which depends, among other factors, on the evolutionary phase
of the initial subgiant, the amount of mass that has been transferred
to the companion before the supernova and the amount of mass stripped
by the supernova impact itself. To illustrate this sensitivity we
performed a similar series of thermal equilibration calculations for
the same initial subgiant but where we stripped off the outermost
0.8\Ms, and heated the remaining outer 83\,per cent in radius
containing 0.03\,\Ms, by various amounts of energy from $5\times
10^{45}\,$erg to $3.5\times 10^{46}\,$erg; the latter is close
to the energy needed to unbind the envelope in this case ($\sim
3.7\times 10^{46}\,$erg). Because of the lower remaining mass, the star
evolves much more rapidly.  In Figure~2 the shaded (yellow) region
shows the location of the secondary in the H-R diagram between $10^3$
and $10^4$\,yr in these calculations. Note that, in all cases, the
secondary is significantly underluminous (by up to a factor of 10) --
even in the strongly heated case -- and may have the appearance of a
main-sequence star similar to the Sun (but with much lower gravity).

\section{Discussion and Conclusions}

In this paper we have examined the evolution of a typical subgiant
secondary (as expected in one of the most popular progenitor models
for SNe Ia) that has been heated and stripped by the impact of a
supernova shell. We have demonstrated that the secondary may be
significantly underluminous or overluminous at different times during
the post-impact phase when it tries to re-establish thermal
equilibrium.  This depends on the amount of heating it has experienced
and the amount of envelope mass that has been stripped off by the
impact or has been lost by mass transfer before the explosion.  Based on
our calculations, we typically expect its luminosity $10^3\,$yr after
the impact to be within an order magnitude of the pre-supernova
luminosity (either above or below). If the outer envelope cools
dramatically (which happens if most of the internal luminosity goes
into driving internal expansion), the secondary may become extremely
underluminous. While it would be quite faint, it would be a rather
unusual, extremely red and probably highly variable subgiant that
should stand out because of its anomalous appearance.  Even if it has
the appearance of a more typical subgiant, where it may in fact be
somewhat hotter (more yellow) than a standard subgiant, one may be
able to see changes in its appearance on a human timescale because of
the short thermal timescale of the outermost layers (which may be
many orders of magnitude shorter than the Kelvin-Helmholtz
timescale of the pre-supernova star).

The main implication for searches of surviving companion stars in
historical supernova remnants is that, in order to be able to rule out
a subgiant companion with any degree of confidence, one has to go to
much fainter magnitudes than previously assumed (down to a luminosity
of $\sim 10^{-1}\Ls$ for a $10^3\,$yr-old remnant, a luminosity of
$\sim 10^{-2}\Ls$ for a $10^4\,$yr-old one). We still expect that it
will have an appearance and/or show signs of variability that makes
it easily identifiable as the supernova companion (in addition to its
runaway space velocity and possible chemical anomalies).

If no plausible candidate for a subgiant companion is found at this
level or for any other type of pre-supernova companion, this would give
some weight to other, presently less favoured progenitor models where
no companion is expected (e.g. the double-degenerate merger model) or
where the companion may already have become a very cool white dwarf
(provided that there is a long time delay between the white-dwarf
accretion phase and the supernova, as expected in some models).

\section*{Acknowledgements}

This work was initiated during a workshop on Type Ia supernovae in
July 2002 at the Lorentz Center in Leiden organized by B. Schmidt and
N. Langer. The author thanks the organizers and many other
participants for numerous fruitful discussions and thanks the Center
for its support.  This work was also in part supported by a European
Research and Training Network Grant on Type Ia Supernovae
(HPRN-CT-20002-00303).

\label{lastpage}

\end{document}